\documentclass[runningheads]{llncs}
\usepackage{graphicx}
\usepackage{algorithm} 
\usepackage{amssymb}
\usepackage{amsmath}
\usepackage{algorithmic}
\usepackage{indentfirst}
\usepackage{subfigure}
\usepackage{multirow}
\usepackage{color}
\usepackage{appendix}
\begin{document}
\title{Content-Preserving Diffusion Model for Unsupervised AS-OCT image Despeckling}
%

\author{Sanqian Li\inst{1} \and
Risa Higashita\inst{1,2*} \and
Huazhu Fu \inst{3}\and
Heng Li \inst{1} \and
Jingxuan Niu \inst{1} \and
Jiang Liu\inst{1*}}
\authorrunning{S. Li et al.}
%
\institute{Research Institute of Trustworthy Autonomous Systems and Department of Computer Science and Engineering, Southern University of Science and Technology, Shenzhen, China \and
Tomey Corporation, Japan \and
Institute of High Performance Computing (IHPC), Agency for Science, Technology and Research (A*STAR), Singapore\\
\email{\{risa,liuj\}@mail.sustech.edu.cn}}

\maketitle              
\begin{abstract}
Anterior segment optical coherence tomography (AS-OCT) is a non-invasive imaging technique that is highly valuable for ophthalmic diagnosis. However, speckles in AS-OCT images can often degrade the image quality and affect clinical analysis. As a result, removing speckles in AS-OCT images can greatly benefit automatic ophthalmology analysis. Unfortunately, challenges still exist in deploying effective AS-OCT image denoising algorithms, including collecting sufficient paired training data and the requirement to preserve consistent content in medical images.
To address these practical issues, we propose an unsupervised AS-OCT despeckling algorithm via Content Preserving Diffusion Model (CPDM) with statistical knowledge.
At the training stage, a Markov chain transforms clean images to white Gaussian noise by repeatedly adding random noise and removes the predicted noise in a reverse procedure. At the inference stage, we first analyze the statistical distribution of speckles and convert it into a Gaussian distribution, aiming to match the fast truncated reverse diffusion process. We then explore the posterior distribution of observed images as a fidelity term to ensure content consistency in the iterative procedure.
Our experimental results show that CPDM significantly improves image quality compared to competitive methods. Furthermore, we validate the benefits of CPDM for subsequent clinical analysis, including ciliary muscle (CM) segmentation and scleral spur (SS) localization.

\end{abstract}
\section{Introduction}

Anterior segment optical coherence tomography (AS-OCT) is a widely used non-invasive imaging modality for ocular disease~\cite{SunitaRadhakrishnan2001RealTimeOC,Leung2011}. It produces high-resolution views of superficial anterior segment structures, such as the cornea, iris, and ciliary body. However, speckle noise inherently exists in AS-OCT imaging systems~\cite{JosephMSchmitt1999SpeckleIO}, which can introduce uncertainty in clinical observations and increase the risk of misdiagnosis. AS-OCT despeckling has become an urgent pre-processing task that can benefit clinical studies.

To suppress speckle noise in AS-OCT images, commercial scanners~\cite{MahnooshTajmirriahi2021ModelingOR} generally average repeated scans at the same location. However, this approach can result in artifacts due to uncontrollable movement. As a result, several post-processing denoising approaches have been developed to reduce speckles, such as wavelet-modified block-matching and 3D filters~\cite{BoChong2013SpeckleRI}, anisotropic non-local means filters~\cite{JaehongAum2015EffectiveSN}, and complex wavelets combined with the K-SVD method~\cite{RahelehKafieh2015ThreeDD}. However, these algorithms can lead to edge distortion depending on the aggregation of similar patches. 
Deep learning has recently been employed for image despeckling tasks, with promising performance~\cite{YuhuiMa2018SpeckleNR}. To overcome the limitations caused by the requirement for vast supervised paired data, unsupervised algorithms explore some promising stages to loosen the paired clinical data collection, including cycle consistency loss~\cite{JunYanZhu2017UnpairedIT}, contrast learning strategies~\cite{TaesungPark2020ContrastiveLF}, simulated schemes~\cite{RudigerGobl2022Speckle2SpeckleUL}, or the Bayesian model~\cite{AndreaBordoneMolini2020Speckle2VoidDS}.
Alternatively, the denoising diffusion probabilistic model (DDPM) can use the averaged image of repeated collections to train the model with excellent performance due to its focus on the noise pattern rather than the signal~\cite{JonathanHo2020DenoisingDP}. Given the prominent pixel-level representational ability for low-level tasks, diffusion models have also been introduced to medical image denoising based on the Gaussian assumption of the noise pattern~\cite{HyungjinChung2022MRID,DeweiHu2022UnsupervisedDO}. 

Although previous studies have achieved outstanding performances, deploying AS-OCT despeckling algorithms remains challenging due to several reasons: \textbf{(1).} Gathering massive paired data for supervised learning is difficult because clinical data acquisition is time-consuming and expensive. \textbf{(2).} Speckle noise in AS-OCT images strongly correlates with the real signal, making the additive Gaussian assumption on the speckle pattern to remove noise impractically. \textbf{(3).} unsupervised algorithms can easily miss inherent content, and content consistency are vital for clinical scenes. \textbf{(4).} Existing algorithms focus on suppressing speckles while ignoring the performance improvement of clinical analysis from despeckling results. 

To address these challenges, we propose a Content-Preserving Diffusion Model for AS-OCT despeckling, named \textbf{CPDM}, which removes speckle noise in AS-OCT images while preserving the inherent content simultaneously. \textbf{Firstly}, we efficiently remove noise via a conditioned noise predictor by truncated diffusion model~\cite{HyungjinChung2022MRID} in the absence of supervised data. We convert the speckle noise into an additive Gaussian pattern by considering the statistical distribution of speckles in AS-OCT to adapt to the reverse diffusion procedure. 
\textbf{Secondly}, we incorporate the posterior probability distribution in observed AS-OCT images into an iterative reverse stage to avoid getting trapped in artificial artifacts and preserve consistent content. The posterior distribution is regarded as a data fidelity term to constrain the iterative reverse procedure for despeckling. \textbf{Finally}, experiments on the AS-Casia and CM-Casia datasets demonstrate the effectiveness of CPDM compared to state-of-the-art (SOTA) algorithms. Further experiments on ciliary muscle (CM) segmentation and scleral spur (SS) localization verify that the CPDM can benefit clinical analysis.

\section{The statistical characteristic of speckles} 

 \begin{figure}[!t]
\centering
\centerline{\includegraphics[width=8cm]{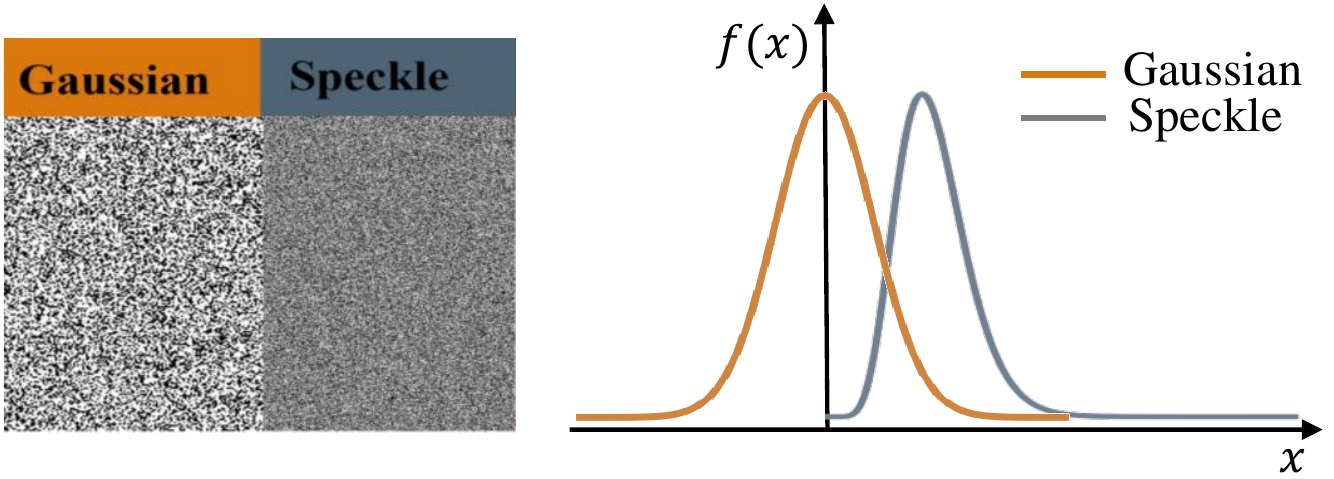}}
\caption{Distributions of Gaussian and speckle noises.} 
\label{Figure:0}
\end{figure}

Speckle noise is inherent in coherent imaging systems~\cite{JosephMSchmitt1999SpeckleIO}, as it results from the destructive interference of multiple-scattered waves.
As shown in Fig.~\ref{Figure:0}, unlike the additive Gaussian noise~${{Y}_{i}}={{x}_{i}}+{{N}_{i}}~(i=1,...,n)$, the multiplicative speckle noise is modeled as ${{Y}_{i}}={{x}_{i}}{{N}_{i}}$~\cite{ZahraAmini2016StatisticalMO}, where $Y$ denotes the noisy image, $x$ is the noise-free image, $N$ is the speckle noise, and $i$ is the pixel index. Moreover, $N$ consists of independent and identically distributed random variables with unit mean, following a gamma probability density function ${{p}_{N}}$~\cite{JoseMBioucasDias2009MultiplicativeNR}:
\begin{equation}
     {{p}_{N}}(n)=\frac{{{M}^{M}}}{\Gamma (M)}{{n}^{M-1}}{{e}^{-nM}},
\label{Eq:1}
\end{equation}
where $\Gamma (\cdot )$ is the Gamma function and $M$ is the number of multilook~\cite{JoseMBioucasDias2009MultiplicativeNR}. To transform the multiplicative noise into an additive one, logarithmic transform~\cite{MohamadForouzanfar2007ADM} is employed on both sides of Eq.~\ref{Eq:1}, as: $\underbrace{\log Y}_{G}=\underbrace{\log x}_{z}+\underbrace{\log N}_{W}$. Therefore, the density of the random variable $W=\log N$ is $ {{p}_{W}}(w)={{p}_{N}}({{e}^{w}}){{e}^{w}}=\frac{{{M}^{M}}}{\Gamma (M)}{{e}^{Mw}}{{e}^{-{{e}^{w}}M}}$. According to the central limit theorem and analyzing the statistical distribution of transformed one in~\cite{boyer2006automatic,dubose2017statistical}, $W$ approximately follows a Gaussian distribution. 
Besides, we can obtain the prior distribution:
\begin{equation}
 {{p}_{G\left| z \right.}}(g\left| z \right.)=pW(g-z).
\label{Eq:2}
\end{equation}

\section{Content Preserving Diffusion Model} 

\begin{figure}[!t]
\centering
\centerline{\includegraphics[width=12cm]{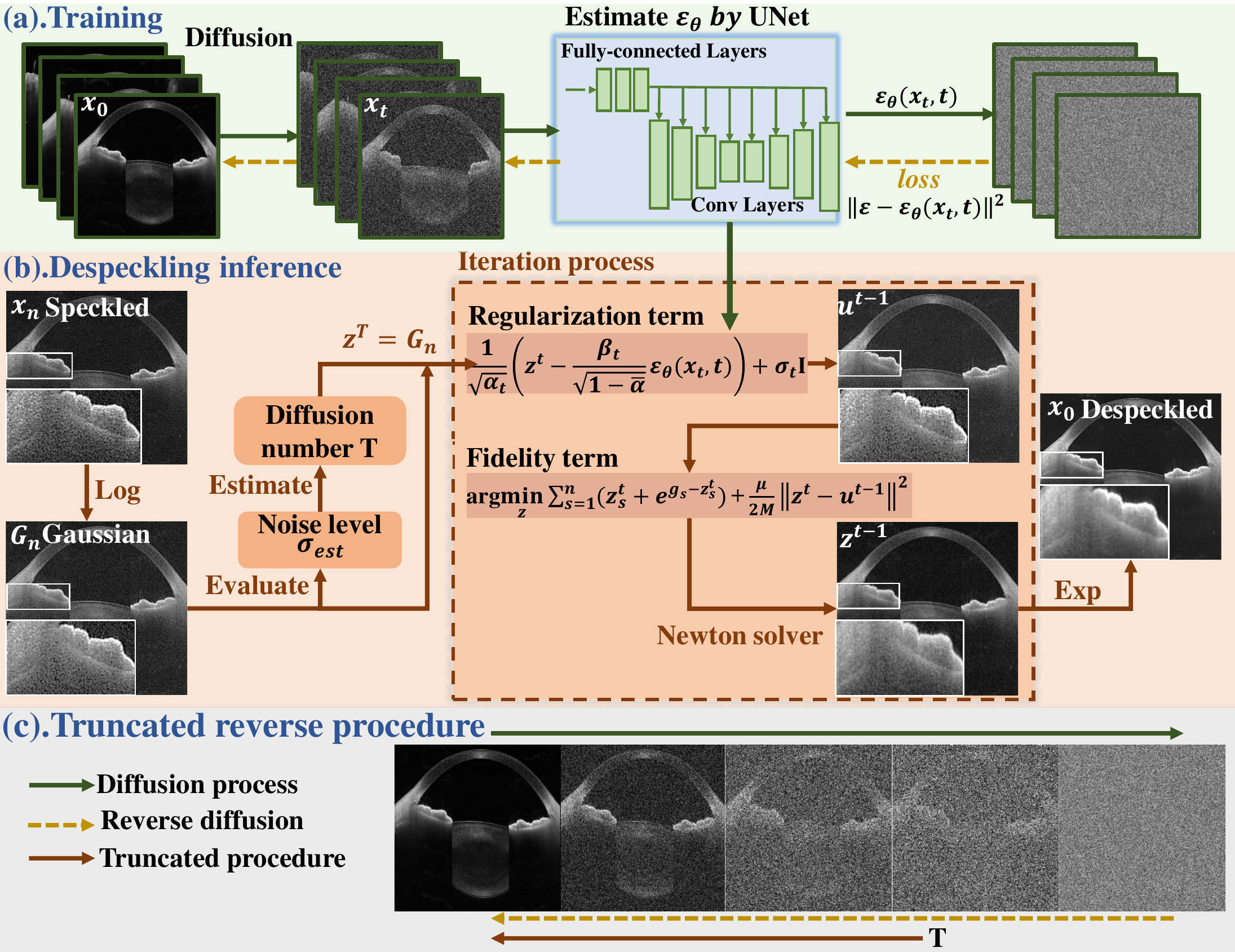}}
\caption{Illustration of proposed CPDM algorithm. CPDM follows the training network in block(a), and learns the regularization knowledge from the trained network for image despeckling shown in block(b). Moreover, we adopt the truncated strategy shown in block(c) into the despeckling process.} 
\label{Figure:1}
\end{figure}

\textbf{Diffusion model}. The diffusion model can subtly capture the semantic knowledge of the input image and prevails in the pixel-level representation~\cite{JonathanHo2020DenoisingDP}. As shown in Fig.~\ref{Figure:1}.(a), it defines a Markov chain that transforms an image ${{x}_{0}}$ to white Gaussian noise ${{x}_{T}}\sim \mathcal{N}(0,1)$ by adding random noise in $T$ steps. During inference, a random noise ${{x}_{T}}$ is sampled and gradually denoised until it reaches the desired image ${{x}_{0}}$. To perfectly recover the image in the reverse sampling procedure, a practicable constraint ${{D}_{KL}}(q({{x}_{t-1}}\left| {{\text{x}}_{t}} \right.,{{x}_{0}})\left\| {{p}_{\theta }} \right.({{x}_{t-1}}{{\left| \text{x} \right.}_{t}}))$ was proposed to minimize the distance between  
$\text{ }{{p}_{\theta }}({{x}_{t-1}}\left| {{\text{x}}_{t}} \right.)$ and
$\text{ }q({{x}_{t}}\left| {{\text{x}}_{t-1}} \right.)$\cite{JonathanHo2020DenoisingDP}. Thus ${x}_{t-1}$ can be sampled as follows:
\begin{equation}
\text{x}_{t-1}=\frac{1}{\sqrt{\alpha_t}}({x_t}-\frac{{\beta_t}}{\sqrt{1-{\bar\alpha_t}}}{\varepsilon_\theta}({x_t},t))+\sigma_t I, 
\label{Eq:3}
\end{equation}
where ${{\varepsilon }_{\theta }}$ is an approximator intended to predict noise ${\varepsilon }$ from ${{\text{x}}_{t}}$ and $I\sim \mathcal{N}\text{(0,1)}$. 

\textbf{Truncated diffusion model}. As mentioned in the previous section, speckle noise follows a gamma distribution and can be transformed into a Gaussian distribution via a logarithmic function. This transformation enables matching the Markov chain procedure in the reverse diffusion process. To speed up the sampling process, this work introduces a truncated reverse procedure that can directly obtain satisfying results from posterior sampling~\cite{HyungjinChung2022MRID}. Fig.~\ref{Figure:1}.(c) illustrates that only the last few reverse diffusion iterations calculated by parameter estimation technique~\cite{GuangyongChen2015AnES} are used to obtain the desired result during despeckling inference. Specifically, following~\cite{JonathanHo2020DenoisingDP}, a Markov chain adds Gaussian noise to the data until it becomes pure noise and then gradually removes it by the reverse procedure at the training stage shown in Fig.~\ref{Figure:1}.(a). At the despeckling inference stage shown in Fig.~\ref{Figure:1}.(b), speckled images are converted into additive Gaussian ones by applying a logarithmic function. Then, the iteration number is determined by estimating the noise levels~\cite{HyungjinChung2022MRID} to achieve an efficient and effective truncated reverse diffusion procedure. Therefore, AS-OCT despeckling can start from noisy image distributions rather than pure noise.

\textbf{CPDM integrated fidelity term}. Inspired by the fact that the score-based reverse diffusion process is a stochastic contraction mapping so that as long as the data consistency imposing mapping is non-expansive, data consistency incorporated into the reverse diffusion results in a stochastic contraction to a fixed point~\cite{HyungjinChung2023ComeCloserDiffuseFasterAC}. This work adopts the theory into the inverse AS-OCT image despeckling problems, as the iteration steps which impose fidelity term can be easily cast as non-expansive mapping. Accordingly, we can design a fidelity term to achieve data consistency by modeling image despeckling inverse problem. Specifically, invoking the conditional independence assumption, the prior distribution with Eq.~\ref{Eq:2} can be rewritten as:
\begin{equation}
    \log {{p}_{G\left| z \right.}}(g\left| z \right.)=\sum\limits_{s=1}^{n}{\log pW({{g}_{s}}-{{z}_{s}})}=C-M\sum\limits_{s=1}^{n}{({{z}_{s}}+{{e}^{{{g}_{s}}-{{z}_{s}}}})}.
\label{Eq:4}
\end{equation}
The Bayesian maximum a posteriori (MAP) formulation leads to the image despeckling optimization with data fidelity and regularization terms.  
\begin{equation}
     \arg \mathop {\min }\limits_z M\sum\limits_{s = 1}^n {({z_s} + {e^{{g_s} - {z_s}}})}  + \lambda R(z),
\label{Eq:5}
\end{equation}
where $R()$ is the regularization term, and $\lambda$ is the regularization parameter.
The unconstrained minimization optimization problem can be defined as a constrained formulation by variable splitting method \cite{MagnusRHestenes1969MultiplierAG}:
\begin{equation}
    (\hat z,\hat u) = arg\mathop {min}\limits_{z,{\rm{u}}} M\sum\limits_{s = 1}^n {({z_s} + {e^{{g_s} - {z_s}}})}  + \lambda R(u)~~{\rm{  s}}{\rm{.t}}{\rm{.  }} ~~ z = {\rm{u}}.
\label{Eq:6}
\end{equation}
 Motivated by the iterative restoration methods with prior information to tackle various tasks become mainstream,  we explore the fidelity term Eq.~\ref{Eq:4} from the posterior distribution of observed images into the iterative reverse diffusion procedure. The fidelity can guarantee data consistency with original images and avoid falling into artificial artifacts. Moreover, we learn reasonable prior from DDPM reverse recover procedure, which can ensure the flexibility with iterative fidelity term incorporated into the loop of prior generation procedure. As shown in Fig.~\ref{Figure:1}.(b), the recovery result obtained from the reverse sampling of DDPM (Eq.~\ref{Eq:3}) can be considered as regularization information of the image despeckling optimization model, and the fidelity term in Eq.~\ref{Eq:5} can ensure the consistency of the reverse diffusion process with the original image content. Therefore, we can achieve AS-OCT image despeckling by solving Eq.~\ref{Eq:6} with the ADMM method using variable splitting technique~\cite{JoseMBioucasDias2009MultiplicativeNR}:
\begin{align} 
    {{\text{u}}^{t-1}} &=\frac{1}{\sqrt{{{\alpha }_{t}}}}({{z}^{t+1}}-\frac{{{\beta }_{t}}}{\sqrt{1-{{{\bar{\alpha }}}_{t}}}}{{\varepsilon }_{\theta }}({{z}^{t+1}},t))+{{\sigma }_{t}}I,
\label{Eq:7} \\
    {z^{t-1}} & \leftarrow \arg \mathop {\min }\limits_z \sum\limits_{s = 1}^n {(z_s^{t} + {e^{{g_s} - z_s^{t}}})}  + \frac{\mu }{{2M}}{\left\| {{z^{t}} - {u^{t-1}}} \right\|^2},
\label{Eq:8}
\end{align} 
where the hyperparameter $u$ control the degree of freedom. It is worth mentioning that Eq.~\ref{Eq:7} is obtained with the trained CPDM model, and Eq.~\ref{Eq:8} can be solved by the Newton method~\cite{SEHMvanBree2020ANS}. Finally, we design an AS-OCT image despeckling scheme by adopting a fidelity term integrated statistical priors to preserve content in the iterative reverse procedure.

\section{Experiment}
To evaluate the performance of the proposed CPDM for AS-OCT image despeckling, we conduct the comparative experiment and a ablation study in three evaluations, including despeckling evaluation, subsequent CM segmentation or SS localization.

\textbf{Dataset preparation}. A series of unsupervised methods including generative adversarial networks (GAN) and diffusion models aim at learning the noise distribution rather than the signal. Therefore, we collect images by averaging 16 repeated B-scans as noisy-free data collected from AS-OCT, the CASIA2 (Tomey, Japan). This study obeyed the tenets of the Declaration of Helsinki and was approved by the local ethics committee.

\textbf{\textit{AS-Casia}} dataset contains 432 noisy image and 400 unpaired clean image with the size of $2131\times1600$, which are views of the AS structure, including lens, cornea, and iris. 400 noisy data and 400 clean images were used for training, and the rest were for testing. The SS location in the noisy image is annotated by ophthalmologists.

\textbf{\textit{CM-Casia}} dataset consists of 184 noisy images and 184 unpaired clean data with the size of $1065\times1465$ that show the scope of CM tissue. 160 noisy images and 160 clean data are utilized for training network, with the remaining images reserved for testing. Moreover,  ophthalmologists annotated the CM regions on the noisy images. 

\begin{table}[!t]\scriptsize
\setlength{\abovecaptionskip}{0.cm}
\setlength{\belowcaptionskip}{0.cm}
\centering
\caption{Quantitative evaluation of different methods.}
\label{Table:1}
\setlength{\tabcolsep}{0.8mm}{
\begin{tabular}{c|ccc|c|ccc|cc} 
\hline
 \multicolumn{1}{c|}{Dataset} &\multicolumn{4}{c|}{AS-Casia} &\multicolumn{5}{c}{CM-Casia}\\  
\hline
 \multicolumn{1}{c|}{Task} &\multicolumn{3}{c|}{Despeckling } &\multicolumn{1}{c|}{Localization }&\multicolumn{3}{c|}{Despeckling }&\multicolumn{2}{c}{Segmentation}\\ 
\hline
Method     & CNR$\uparrow$     & ENL$\uparrow$   & NIQE$\downarrow$  &ED$\downarrow$ (um)   & CNR$\uparrow$     & ENL$\uparrow$    & NIQE$\downarrow$  & F1$\uparrow$   & IoU$\uparrow$  \\ 
\hline
Noisy   & 0.52  & 5.12  & 7.05 &57.09  &-6.66 &2.50 &11.50  & 0.579  & 0.424      \\
WBM3D\cite{BoChong2013SpeckleRI}      & 1.15  & 6.74  & 6.31  &56.57  &-3.25 &3.28 &6.80 & 0.602  & 0.447      \\
NLM\cite{AntoniBuades2005ANA}       & 1.76  & 22.94 & 6.54  &96.85  &-0.54 &42.37 &7.39  & 0.657  & 0.508      \\
ANLM\cite{JaehongAum2015EffectiveSN}      & 1.64  & 10.14 & 6.63 &91.97  &-2.18 &4.18 &6.52  & 0.627  & 0.474      \\
WKSVD\cite{RahelehKafieh2015ThreeDD}      & 1.05  & 6.70  & 7.94  &79.60  &-4.98 &5.36 &8.33  & 0.681  & 0.531      \\
UINT\cite{MingYuLiu2017UnsupervisedIT}      & 2.14  & 6.45  & 9.04 &121.98   &-1.60 &12.98 &9.03  & 0.641  & 0.492      \\
CUT\cite{TaesungPark2020ContrastiveLF}      & 1.94  & 5.47  & 6.23  &83.05  &-4.61 &5.99 &6.92  & 0.553  & 0.404      \\
CycleGAN\cite{JunYanZhu2017UnpairedIT} & 1.82  & 5.13  & 5.58 &65.42  &-3.12 &11.17 &7.44 & 0.667  & 0.516      \\
Speckle2void\cite{AndreaBordoneMolini2020Speckle2VoidDS} & 0.51  & 5.07  & 7.06 &59.79  &-5.47  &4.71 &7.86  & 0.665 & 0.514      \\
DRDM\cite{HyungjinChung2022MRID} &1.28  &21.23  &5.86 &37.96  &-4.98  &33.09&9.18  & 0.670 & 0.524      \\
\hline
ODDM\cite{DeweiHu2022UnsupervisedDO}   & 0.14  & 31.33  & 6.11  &38.18  &-7.06 &91.50 &9.70   & 0.330  & 0.224      \\
LogDM   & 1.63  & 21.16  & 5.27 &38.04 &-2.08 &139.25 &7.70  & 0.679  & 0.535      \\
\textbf{CPDM}
  & \textbf{2.16}  & \textbf{143.68}  
&\textbf{4.84} &\textbf{37.43} &\textbf{-0.53} & \textbf{396.35} & \textbf{6.42}  & \textbf{0.703}  & \textbf{0.561}      \\
\hline
\multicolumn{1}{l}{}  & \multicolumn{1}{l}{} & \multicolumn{1}{l}{} & \multicolumn{1}{l}{} & \multicolumn{1}{l}{} & \multicolumn{1}{l}{} & \multicolumn{1}{l}{} 
\end{tabular}}
\end{table}

\textbf{Implementation settings}. The backbone of our model is a simplified version of that in~\cite{JonathanHo2020DenoisingDP}. The CPDM network was trained on an NVIDIA RTX 2080TI 48GB GPU for 500 epochs, with a batch size of 2, using Adam optimizer. The variance schedule is set to linearly increase from ${{10}^{-4}}$ to ${{6}^{-3}}$ in $T=1000$ steps and the starting learning rate is ${{10}^{-4}}$ and decay by half every 5 epochs. All training images were resized to $512\times512$ and normalized to [0,1]. In the reverse procedure, we analyzed the hyperparameters and set $\lambda =\mu /2M=0.2$, and the iteration number was evaluated as $T=4$. The parameters of classical blind denoising methods were tuned to reach the best performance, in which the noise-level of NLM, ANLM, WBM3D, and WK-SVD were conducted by~\cite{GuangyongChen2015AnES}. The recent unsupervised methods, UINT, CUT, and CycleGAN are conducted by the unpaired dataset with the default setting. The methods based on diffusion models only are trained on clean data while Speckle2void is implemented only on noisy data with the default setting.

\begin{figure}[!t]
\centering
\centerline{\includegraphics[width=1\linewidth]{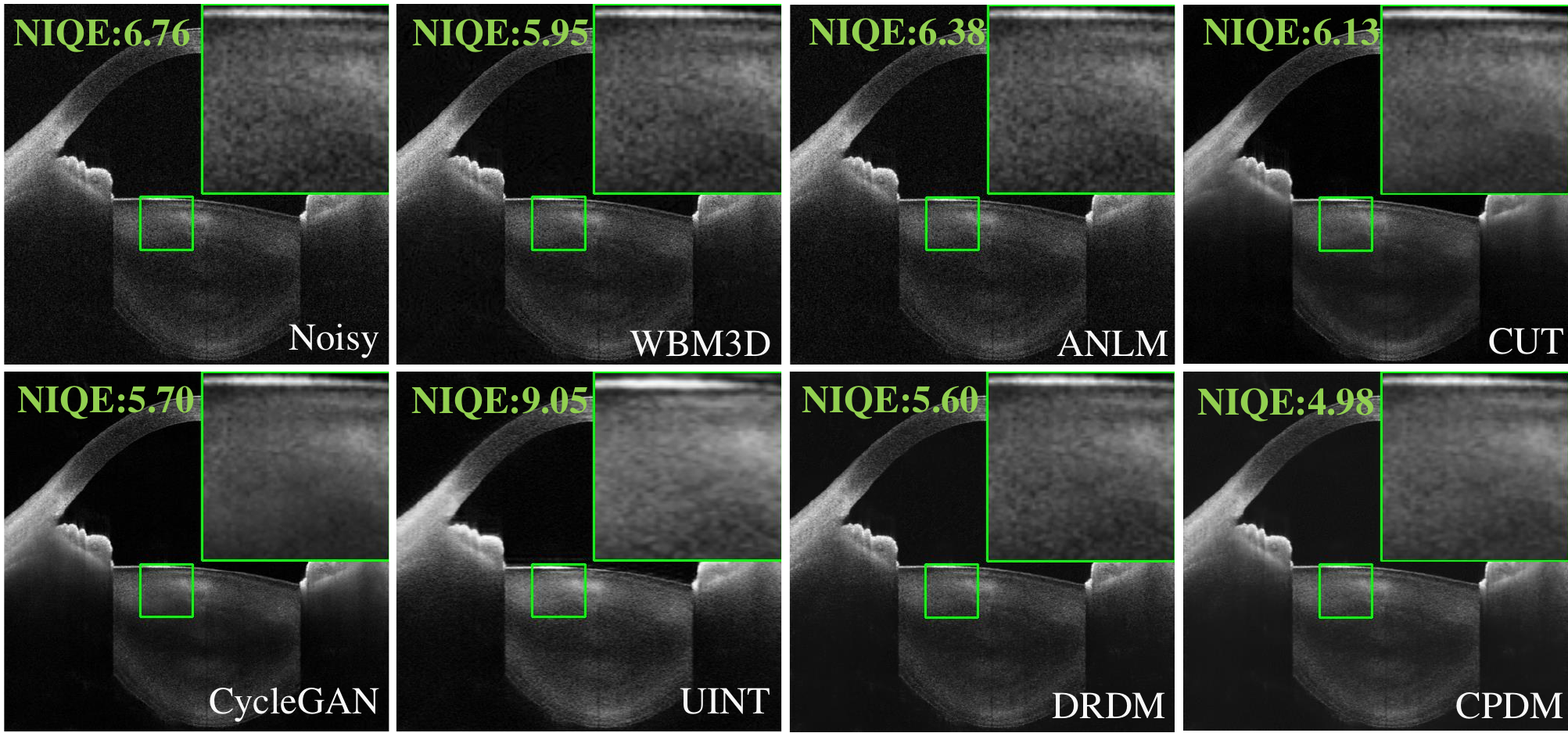}}
\caption{The visual comparison of image despeckling results} 
\label{Figure:3}
\end{figure}

\textbf{Comparison on AS-Casia dataset}. We first evaluate the despeckling performance by parameterless index, including contrast-to-noise ratio (CNR)~\cite{YuhuiMa2018SpeckleNR}, the equivalent number of looks (ENL)~\cite{YuhuiMa2018SpeckleNR}, and natural image quality evaluator (NIQE)~\cite{mittal2012making}. Then we compare the despeckling results with the SOTA methods by using the SS localization task with trained models in~\cite{PengLiu2022ReproducibilityOD}. Concretely, we calculate a euclidean distance (ED) value between the reference and the predicted SS position with despeckled images via trained models. As shown in Table~\ref{Table:1}, the proposed CPDM achieves promising despeckling results in terms of the best CNR, ENL, NIQE values and the minimum ED error in the SS localization task among all approaches. The visual comparison for denoised images with competing approaches is shown in Fig.~\ref{Figure:3}: the \textcolor{green}{green} region has been enlarged to highlight the structure of the anterior lens capsule, which can assist in diagnosing congenital cataracts. It can be observed that the CUT and CycleGAN models oversmooths structures close to flat, the UINT method results in ringing effects while the WBM3D, ANLM and DRDM algorithms retain speckles in the lens structure. Obviously, the proposed CPDM acquires satisfactory quality with fine structure details and apparent grain. 

\begin{figure}[!t]
\centering
\centerline{\includegraphics[width=1\linewidth]{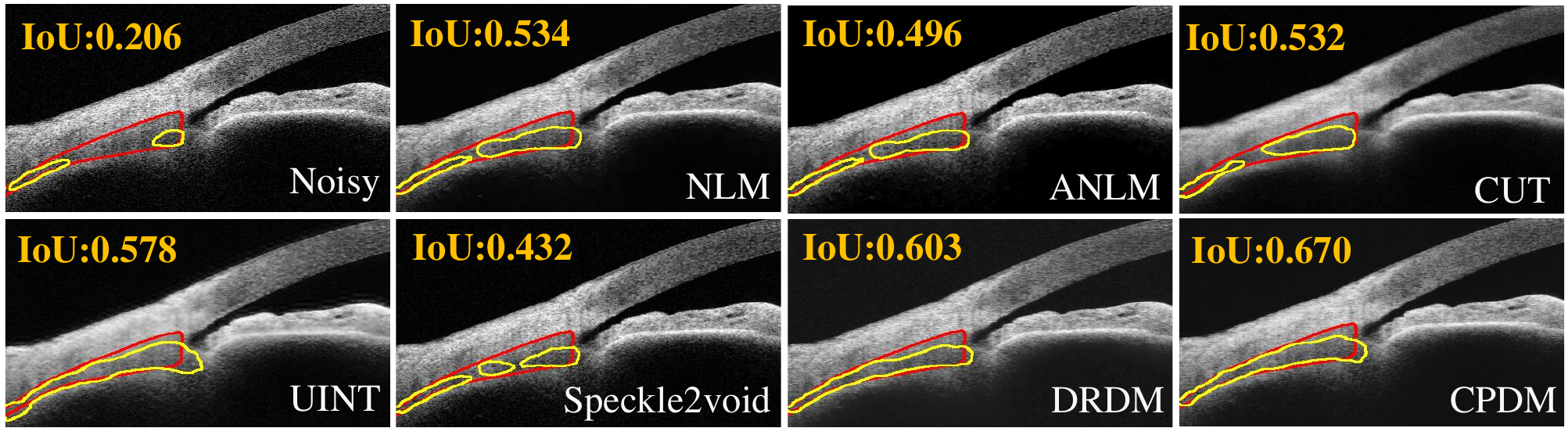}}
\caption{Comparisons of CM segmented results} 
\label{Figure:4}
\end{figure}

\textbf{Comparison on CM-Casia dataset}. We conduct the experiment of image despeckling and the following CM segmentation task to validate the clinical benefit with CPDM. Specifically, we train a U-Net segmentation model~\cite{OlafRonneberger2015UNetCN} on the CM-Casia dataset and then test the despeckled images of various methods. F1-Score and intersection over union (IoU) index for segmentation were calculated between the despeckled images and reference as reported in Table.~\ref{Table:1}. It can be seen that the proposed CPDM achieves the superior despeckling performance by the highest CNR, ENL, NIQE values and segmentation metrics. Moreover, the segmented CM example of competitive methods is depicted in Fig.~\ref{Figure:4}, in which the CM boundaries reference with the \textcolor{red}{red} line, and the \textcolor{yellow}{yellow} line means the segmented results. We can see that NLM, ANLM, CUT, and Speckle2void methods fail to the continuous segmentation results due to insufficient speckle suppression or excessive content loss while the CPDM captures a distinct CM boundary and obtains the highest IoU score. Notably, as a type of smooth muscle, CM has ambiguous boundaries, which are easily affected by speckles, resulting in difficulty distinguishing CM from the adjacent sclera and negative CNR values. Despite these challenges, the proposed CPDM can achieve the best segmentation owing to the speckle reduction while preserving the inconspicuous edge content. 

\textbf{Ablation study}. Table~\ref{Table:1} shows the ablation study of the proposed CPDM. We compare our method with two variants: ODDM~\cite{DeweiHu2022UnsupervisedDO} and logDM. The ODDM only considers removing the speckles by hijacking the reverse diffusion process with the Gaussian assumption on speckles. Based on the ODDM, the logDM further transforms speckles to Gaussian distribution by analyzing the statistical characteristics of speckles. Additionally, the CPDM adopts the data fidelity term to regulate the despeckling reverse process by integrating content consistency. From Table~\ref{Table:1}, we can see that both the logarithmic function and data fidelity term can improve the quality of despeckled images and benefit the subsequent clinical analysis. Consequently, a prominent unsupervised CPDM to AS-OCT image despeckling is acquired with the proposed strategies.

\section{Conclusions}
Due to the impact of speckles in AS-OCT images, monitoring and analyzing the anterior segment structure is challenging. To improve the quality of AS-OCT images and overcome the difficulty of supervised data acquisition,  we propose a content-preserving diffusion model to achieve unsupervised AS-OCT image despeckling. We first analyze the statistical characteristic of speckles and transform it into Gaussian distribution to match the reverse diffusion procedure. Then the posterior distribution knowledge of AS-OCT image is designed as a fidelity term and incorporated into the iterative despeckling process to guarantee data consistency. Our experiments show that the proposed CPDM can efficiently suppress the speckles and preserve content superior to the competing methods. Furthermore, we validate that the CPDM algorithm can benefit medical image analysis based on subsequent CM segmentation and SS localization task.    

\bibliographystyle{IEEEbib}
\bibliography{refs}

\newpage
\section{Appendix}
\subsection{The proposed CPDM Algorithm}
Algorithm~\ref{alg1:algorithm} and Algorithm~\ref{alg2:algorithm} summarize the whole training and despeckling procedure of CPDM.
\begin{algorithm} 
\caption{Training} 
\label{alg1:algorithm}       
        \ 1: \textbf{Repeat}\\
            \ \hspace*{0.6cm}${{G}_{0}}\sim \log {{x}_{0}}, {{G}_{0}}\sim q({{G}_{0}})$\\
            \  \hspace*{0.6cm}$t\sim \text{Uniform}(\{1,\ldots ,T\}), \varepsilon \sim \mathcal{N}(0,1)$ \\
            \  \hspace*{0.6cm}Take gradient descent step on:
    ${{\nabla }_{\theta }}{{\left\| \varepsilon -{{\varepsilon }_{\theta }}(\sqrt{{{{\bar{\alpha }}}_{t}}}{{G}_{0}}+\sqrt{1-{{{\bar{\alpha }}}_{t}}}\varepsilon ,t) \right\|}^{2}}$\\
        \ 2:\textbf{Until converged}
\end{algorithm}

\vspace{-1.4cm}
\begin{algorithm}
\caption{Reverse despeckling procedure} 
\label{alg2:algorithm}       
    \ 1: ${{G}_{N'}}\sim \log Y, z={{G}_{N'}}, {{\sigma }_{est}}=E({{G}_{N'}}), t'={{\sigma }^{-1}}({{\sigma }_{est}}), N'=t'N$\\
    \ \hspace*{0.1cm}2: \textbf{For} $t=N'-1:0$ \textbf{do}\\
        \hspace*{0.6cm} $I\sim \mathcal{N}(0,1)$\\
        \hspace*{0.6cm} ${{\rm{u}}^{t - 1}} = \frac{1}{{\sqrt {{\alpha _t}} }}({z^t} - \frac{{{\beta _t}}}{{\sqrt {1 - {{\bar \alpha }_t}} }}{\varepsilon _\theta }({z^t},t)) + {\sigma _t}I$\\
        \hspace*{0.6cm} ${z^{t - 1}} = \arg \mathop {\min }\limits_z \sum\limits_{s = 1}^n {(z_s^t + {e^{{g_s} - z_s^t}})}  + \frac{\mu }{{2M}}{\left\| {{z^t} - {u^{t - 1}}} \right\|^2}$\\
    \ \hspace*{0.1cm}3: \textbf{Return} $x={{e}^{z}}$ 
\end{algorithm}  
\subsection{The iterative optimization in reverse despeckling procedure}
The speckled images is generally formulated as:
${{Y}_{i}}={{x}_{i}}{{N}_{i}},~(i=1,...,n)$,  $N$ follows:~${{p}_{N}}(n)=\frac{{{M}^{M}}}{\Gamma (M)}{{n}^{M-1}}{{e}^{-nM}}$.
By taking the logarithmic function:
\begin{equation}
   \underbrace{\log Y}_{G}=\underbrace{\log x}_{z}+\underbrace{\log N}_{W}. 
\end{equation}
 According the random variable transformation rule:
 \begin{equation}
  W=\log N, ~   
 \end{equation}
\begin{equation}
 {{p}_{W}}(w)={{p}_{N}}({{e}^{w}}){{e}^{w}}=\frac{{{M}^{M}}}{\Gamma (M)}{{e}^{Mw}}{{e}^{-{{e}^{w}}M}},  
\end{equation}
\begin{equation}
{{p}_{G\left| z \right.}}(g\left| z \right.)=pW(g-z). 
\end{equation}
Based on the conditional independence rule:
\begin{equation}
 \log {{p}_{G\left| z \right.}}(g\left| z \right.)=\sum\limits_{s=1}^{n}{\log pW({{g}_{s}}-{{z}_{s}})}=C-M\sum\limits_{s=1}^{n}{({{z}_{s}}+{{e}^{{{g}_{s}}-{{z}_{s}}}})}.   
\end{equation}
Based on the MAP criterion, it becomes an unconstrained optimization:
\begin{equation}
\hat{z}=arg\underset{z}{\mathop{min}}\,M\sum\limits_{s=1}^{n}{({{z}_{s}}+{{e}^{{{g}_{s}}-{{z}_{s}}}})}+\lambda R(z).
\end{equation}
Using the variable splitting method to create a new variable:
\begin{equation}
    (\hat{z},\hat{u})=arg\underset{z,\text{u}}{\mathop{min}}\,M\sum\limits_{s=1}^{n}{({{z}_{s}}+{{e}^{{{g}_{s}}-{{z}_{s}}}})}+\lambda R(u)\text{  s}\text{.t}\text{.  }z=u.
\end{equation}
The quadratic proximal point to punish the constraint term:
\begin{equation}
   (\hat{z},\hat{u})=arg\underset{z,\text{u}}{\mathop{min}}\,M\sum\limits_{s=1}^{n}{({{z}_{s}}+{{e}^{{{g}_{s}}-{{z}_{s}}}})}+\frac{\mu }{2}{{\left\| z-u \right\|}^{2}}+R(u).
\end{equation}
The alternating direction method of multipliers (ADMM) algorithm to solve the above formulation by splitting the variable: $u$ is the medium value obtained from the reverse sampling result with DDPM and ${{z}^{t}}$ can be solved by Newton method:
\begin{equation}
{{u}^{t}}=\frac{1}{\sqrt{{{\alpha }_{t}}}}({{z}^{t+1}}-\frac{{{\beta }_{t}}}{\sqrt{1-{{{\bar{\alpha }}}_{t}}}}{{\varepsilon }_{\theta }}({{z}^{t+1}},t))+{{\sigma }_{t}}I, 
\end{equation}
\begin{equation}
{{z}^{t}}\leftarrow \arg \underset{z}{\mathop{\min }}\,\sum\limits_{s=1}^{n}{(z_{s}^{t+1}+{{e}^{{{g}_{s}}-z_{s}^{t+1}}})}+\frac{\mu }{2M}{{\left\| {{z}^{t+1}}-{{u}^{t}} \right\|}^{2}},
\end{equation}
where the sub-problem about ${{z}^{t}}$ can be solved by Newton method, in which $\frac{u}{2M}$ is a hyper-parameter and denoted as $\lambda $:\\
\textbf{While} ${{z}_{k+1}}-{{z}_{k}}>0.01$\\
${{z}_{k+1}}={{z}_{k}}-\frac{1-{{e}^{(g-z)}}+\lambda (z-{{u}^{t}})}{{{e}^{(g-z)}}+\lambda }$\\
\textbf{Return} ${{z}^{t}}={{z}_{k+1}}$

\subsection{Experimental details}

\begin{figure}[htb]
\centering
\centerline{\includegraphics[width=12cm]{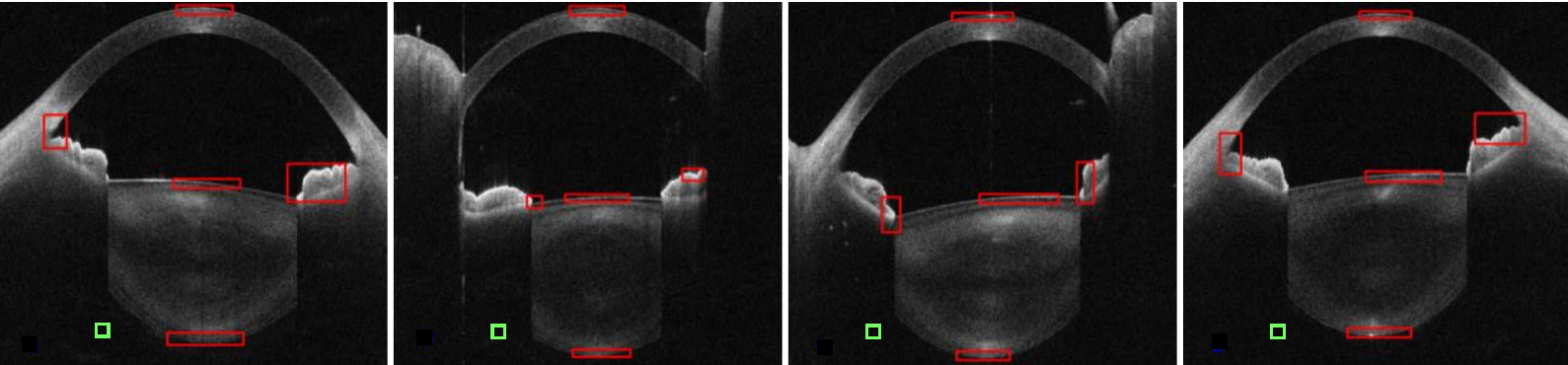}}
\caption{Original noisy examples from AS-casia dataset with selected region of interests (ROIs) and boundaries marked. Five strong contrast signal regions (\textcolor{red}{red}) and one background region (\textcolor{green}{green}) are selected for calculating CNR and ENL.
} 
\label{Figure:7}
\end{figure}

\end{document}